\documentclass[letter]{aa} % for the letters 
\usepackage{graphicx}
\usepackage{txfonts}
\usepackage{lscape}
\begin{document}

   \title{Non-thermal radio emission in Sakurai's Object}

   \author{Marcin Hajduk
          \inst{1}
          \and
          Peter A. M. van Hoof
          \inst{2}
          \and
          Albert A. Zijlstra
          \inst{3}
          \and
          Griet Van de Steene
          \inst{2}
    \and
         Stefan Kimeswenger\inst{4,5} 
         \and \\
         Daniela Barr\'ia \inst{6}
         \and
         Daniel Tafoya \inst{7}
         \and
          Jesús A. Toalá
          \inst{8}
          }

   \institute{Department of Geodesy,
         Institute of Geodesy and Civil Engineering,
         Faculty of Geoengineering,
         University of Warmia and Mazury, ul. Oczapowskiego 2, 10-719 Olsztyn, Poland\\
              \email{marcin.hajduk@uwm.edu.pl}
         \and
             Royal Observatory of Belgium, Ringlaan 3, 1180 Brussels, Belgium
             \and
             Jodrell Bank Centre for Astrophysics, Alan Turing Building, University of Manchester, Manchester M13 9PL, UK        
         \and
         Universit{\"a}t Innsbruck, Institut f{\"u}r Astro- und Teilchenphysik,  Technikerstr. 25/8, 6020 Innsbruck, Austria
         \and
          Universidad Cat\'olica del Norte, Instituto de Astronom{\'i}a,  Av. Angamos 0610, Antofagasta, Chile
          \and
          Universidad Central de Chile, Facultad de Ingenier\'ia y Arquitectura, Av. Francisco de Aguirre 0405, La Serena, Coquimbo, Chile
          \and
          Department of Space, Earth and Environment, Chalmers University of Technology, Onsala Space Observatory, 439 92 Onsala,
Sweden
          \and
          Instituto de Radioastronom\'{i}a y Astrof\'{i}sica, Universidad Nacional Aut\'{o}noma de M\'{e}xico, Morelia 58089, Michoac\'{a}n, Mexico
             }

   \date{Received September 15, 1996; accepted March 16, 1997}

% \abstract{}{}{}{}{} 
% 5 {} token are mandatory
 
  \abstract
   {The very late thermal pulse (VLTP) affects the evolution of $\sim$20\% of 1--8\,$\mathrm M_\odot$ stars, repeating the last phases of the red giant within a few years and leading to the formation of a new, but hydrogen-poor nebula within the old planetary nebula (PN). The strong dust formation in the latter obscures the optical and near-infrared radiation of the star.}
   {We aimed to determine the reheating timescale of the central star in Sakurai's object, which is an important constraint for the poorly understood VLTP evolution.}
   {We observed the radio continuum emission of Sakurai's object for almost 20 years from 2004 to 2023. Continuous, multi-frequency observations proved to be essential to distinguish between phases dominated by photoionization and shock ionization.}
   {The flux density fluctuates by more than a factor 40 within months to years. The spectral index remained negative between 2006 and 2017 and is close to zero since 2019. The emission region is barely resolved since 2021.}
   {Non-thermal radio emission observed from 2004 to 2017 traces shocks induced by wind interactions due to discrete mass-loss events. Thermal emission dominates during the period 2019--2023 and may indicate photoionization of the nebula by the central star.}


   \maketitle
%
%-------------------------------------------------------------------

\section{Introduction}

Low- and intermediate-mass stars ($\rm 1  \, M_{\odot} < M_{\star} < 8 \,
M_{\odot}$) play a vital role in the chemical evolution of the Galaxy. Up to
90\% of their initial mass is returned to the interstellar medium along
with the newly formed products of nucleosynthesis, most of it during the
asymptotic giant branch (AGB) phase of their evolution. AGB stars are
the main source of the $s$-process elements and interstellar carbon dust, and
their contribution of C, N, and F is at least comparable to those of supernovae \citep{2011MNRAS.414.3231K}. 

The main site of carbon production at the end of the AGB phase is the helium-burning shell, which is separated from the hydrogen burning shell by a small inter-shell region. Helium burning is activated in a thermal pulse (helium flash) after the helium has sufficiently accumulated during the rest phases of hydrogen burning. The nuclear synthesized material from the thermal pulse is then transported to the surface by the convective shell. Thermal pulses trigger heavy mass loss due to the increase in luminosity in conjunction with the higher opacity in the convective zone.  When the envelope mass is reduced to $\rm 10^{-3} \, M_{\odot}$, the star %starts evolving
evolves very quickly to higher effective temperatures. It ionizes the ejecta and a planetary nebula (PN) becomes visible. The central star of the PN eventually evolves to a %DA-type 
H-rich white dwarf (WD). 

Stellar evolutionary models anticipated that the last thermal pulse may occur after the extinction of the nuclear burning in the hydrogen shell on the WD cooling track, this is now known as a very late thermal pulse \citep[VLTP;][]{1979A&A....79..108S,1983ARA&A..21..271I}. As a result of the VLTP, the star retraces its evolution to a "born-again" red giant. The remaining hydrogen shell is consumed by the pulse-driven convective zone, leaving a hydrogen-free surface of the star with ${\sim}50\%$ of C and up to 50\% of He by mass, reflecting the intershell abundances. Similar abundances are observed in [Wolf-Rayet] ([WR]) stars (central stars of PNe %which 
whose spectra mimic the population I Wolf-Rayet stars), which suggests that some of these objects might have formed via VLTP \citep[see][]{Todt2015}. The mass ejected due to the VLTP enriches the interstellar medium mainly with helium and carbon and is the main site for the production of carbon dust \citep{2021MNRAS.503.1543T} and some specific isotopes in the Universe, including $\rm ^{13}C$ \citep{2005Sci...308..231H}. Traces of VLTP ejecta were identified in primitive meteorites \citep{2001ApJ...559..463A}.

Only two objects have been observed so far during this very short evolutionary period: Sakurai's Object \citep{1996ApJ...468L.111D} and V605\,Aql \citep{1973ApJ...181..147W}. Other PNe may have experienced a VLTP in the past. The best known examples are A\,30, A\,78 \citep[and references therein]{2021MNRAS.503.1543T}, HuBi\,1 \citep{2018NatAs...2..784G}, and WR\,72 \citep{2020MNRAS.492.3316G}. In this paper we focus on Sakurai's Object, the first well-observed example of a VLTP.

%\textcolor{blue}{JT: We need to introduce here our target, the Sakurai's object.}

\section{Evolution of Sakurai's object}

Sakurai's object (V4334\,Sgr) erupted sometime between 1990 and 1996, as the central star of a faint, old (and previously unknown) PN \citep{1999MNRAS.304..127P}. Optical observations revealed the cooling and expanding photosphere of the star \citep{1997AJ....114.1657D}. This was followed by hydrogen depletion and synthesis of s-process elements and lithium in the atmosphere \citep{1999A&A...343..507A}. The star eventually disappeared from view in the optical in a R\,CrB-like event. Infrared observations revealed the formation of hot dust and molecules \citep{1998MNRAS.298L..37E}. Since then, the source has been monitored in the infrared by several groups \citep[see e.g.][and references therein]{2020ApJ...904...34H,2022MNRAS.511..713E}. 

Soon after the star fully vanished in the optical, emission lines showed up in 2001 \citep{2002ApJ...581L..39K}. The lines declined for a few years after first detection, which was consistent with a cooling shock  \citep{2007A&A...471L...9V}. Since 2008 the lines are continuously brightening \citep{2018Galax...6...79V}.

Sakurai’s Object was first observed at radio wavelengths in 1997 and 2002. The observations showed only the old, H-rich, extended (34\,\arcsec\ in diameter) PN around the central star \citep{1998MNRAS.297..905E,2002Ap&SS.279...69E,2008ASPC..391..163H}. \citet{2005Sci...308..231H} observed a faint point source in the center of the old nebula with the Karl G. Jansky Very Large Array (VLA) in New Mexico in 2004. The source was interpreted as free-free emission from the photoionized region, but showed large variability in the following years which suggested shock-emission instead \citep{2007A&A...471L...9V}.

Observations of Sakurai's object implied %the
a born-again timescale of 5--10 yr, one to two orders faster than predicted by %the
evolutionary models \citep{1996ApJ...468L.111D}. 
Different methods, adopting the physics of convective processes as well as numerical schemes led to new faster evolving model predictions \citep{2001ApJ...554L..71H,2006A&A...449..313M}. \citet{2003ApJ...583..913L} predicted the double-loop evolution of Sakurai's Object in the HR diagram, later secured by \citet{2005Sci...308..231H} and \citet{2006A&A...449..313M}. According to the current understanding of this phenomenon, the remaining hydrogen from the envelope is mixed into the upper layers of the helium-burning shell and ignites separately in a Hydrogen Ingestion Flash (HIF). The first loop from the HIF, with an extremely rapid return to high stellar temperatures, is followed by a much slower loop caused by the helium flash. Although these models predict a return to a nearly 100\,kK source by today, current observations do not suggest this very rapid evolution \citep{2018Galax...6...79V}. 
The strong dust formation in the mass loss has hidden the core of Sakurai's object very efficiently and early from direct optical and infrared observations \citep{2002Ap&SS.279..149K}. 

In this work we have looked inside this dusty cocoon at radio wavelengths to infer indirectly the temperature evolution and mass-loss history.

%These radio observations here are intended to close this gap in the information.

%V605\,Aql VLTP outburst is dated to 1919. It was not followed after the decline until \cite{Seitter} discovered a [WC]-type star in the center of an old PN. Sometimes it is referred as a twin of Sakurai's Object \cite{Clayton} and their evolution is expected to have similar timescales. However, some caveats were added by \cite{Wesson}.

%An indication of reheating of the star came from radio observations of the newly ejected nebula. 

\section{Radio observations}

\subsection{VLA observations}

The radio emission from Sakurai's
object has been monitored since 2004 using the VLA (Table\,\ref{tab:configurations}). While we used 3C286 for flux calibration, J1733$-$1304 was used as phase calibrator. 

Observations before 2012 used two 64\,MHz spectral channels and after 2012 64 64\,MHz spectral channels giving an effective bandwidth of 4\,GHz for the X/C-band and 32 spectral channels giving an effective bandwidth of 2\,GHz for the L-band.

As one of the two spectral channels in 2004 and in 2005 generated an artificial source in the phase center, causing a weak interference pattern in the image, they were flagged. We made the image with the remaining spectral channel. The images were cleaned using the Common Astronomy Software Applications for Radio Astronomy \citep[CASA]{2022PASP..134k4501C} using Briggs' weighting.

%\textcolor{green}{Griet: you give no details about what was used in the imaging and for the cleaning proces (eg weights, boxes etc)} 

We fitted a compact source in the centre of the old PN, at the position of the central star with a 2D Gaussian model with CASA. The fitting yielded the peak flux density, integrated flux density and source position of the gaussian. We used peak flux density in the cases where the source remained unresolved and the integrated flux density for configurations where the source was resolved. Peak value was less sensitive to variable background in cases where the synthesized beam was relatively large and the source was contaminated by the old PN. The source is slightly resolved in 2021, 2022, and 2023 in the X-band and in the C band in 2022. Due to the low elevation of the object at the VLA site (max. elevation of 39 degrees), the synthesized beam is elongated in the N--S direction. This affected the size measurements (less resolved in the N--S direction) and in some cases flux measurements (e.g. in the L band in 2017 and 2021 - integrated flux is much higher than peak flux, although the source was unresolved).

An old, extended, optically thin nebula is present in the B configuration at 1.5\,GHz, CnB and DnC configurations. The old nebula is resolved out in other observations and does not affect the measurements of the central source.

%However, the size of the source was larger in 2021, which would be difficult to interpret. A slight extension can be seen south-west to the point source in 2021, which could be noise in the map. This affects the measurement. Thus, we do not consider the measured sizes to be reliable.

\begin{table}
\centering
\caption{Dates, frequency bands and array configuration of VLA radio observations of Sakurai's Object used in this work.}
%List of VLA radio observations of Sakurai's Object used in this paper: dates, frequency bands, and the array configuration. 
\label{tab:configurations}
\begin{tabular}{ccc}
\hline
\hline
\\[-1.5ex]
Date & Band & Configuration  \\
\\[-1.5ex]
\hline
\\[-1.5ex]
%1997-05-28 & C & B->CnB & & & & & AE113 \\
%1998-11-06 & C & CnB & & & & & AE123  \\
%2002-01-03 & C & D & & & & & AE145  \\
%2002-11-29 & C, K  & C & & & & & AH794 \\
2004-02-05 & X & CnB   \\
2005-02-04 & C & BnA   \\
2005-02-06 & X & BnA   \\
2006-06-11 & X & BnA  \\
2006-06-12 & C & BnA->B  \\
%2007-01-28 & X, K, Ku & DnC  \\
2007-02-02 & X, C, L & DnC  \\
%2007-09-04 & X, K & DnC  \\
2007-09-05 & X, C, L & A  \\
2007-10-10 & X, C, L & BnA  \\
2012-10-08 & X & A \\
2012-10-11 & L,C & A \\
%2013-01-12 & L,C,X & A D  \\
2017-09-12/13 & L,C,X & B  \\
2019-09-04 & L,C,X & A  \\
2021-01-09 & L,C,X & A  \\
2022-06-25 & L,C,X & A  \\
2023-07-23 & X & A  \\
\\[-1.5ex]
\hline
\end{tabular}
\\
\end{table}

%van Hoof 2008 A\&A: 2004, 2005, 2006
%van Hoof 2008 ASPS: 2007

%Radio observations from 09.2008 were reported by \citet{2008ASPC..391..155V}.

%2012 and 2013 observations were reported by \citet{2015ASPC..493...95V}.

\subsection{ATCA observations}

Sakurai's Object was observed with the Australia Telescope Compact Array (ATCA) on October 27, 2008. The rest frequencies were centered at 8.64 and 4.80\,GHz. The flux calibrator was $1934-638$ and the phase calibrator was $1752-225$. We estimated the flux density of the phase calibrator of $1752-225$ as $0.288 \pm 0.014 \, \mathrm{Jy}$ at 4.8\,GHz and $0.344 \pm 0.024 \, \mathrm{Jy}$ at 8.46\,GHz. 

\section{Radio evolution of Sakurai's Object}

%8/10\,GHz light curve suggests that three maxima occurred (Fig.\,\ref{Fig:fluxes}). 

We confirm the 2004 detection by \cite{2005Sci...308..231H}, but did not detect the source reported by \citet{2007A&A...471L...9V} in our re-processed 2005 data. Consequently, we can state that the radio flux of Sakurai declined between 2004 and 2005 because an upper limit in 2005 was below the 2004 detection. Hence the first maximum must have occurred between December 2002 and February 2005, but was unfortunately not covered by our observations.

\begin{figure}
\includegraphics[width=\columnwidth]{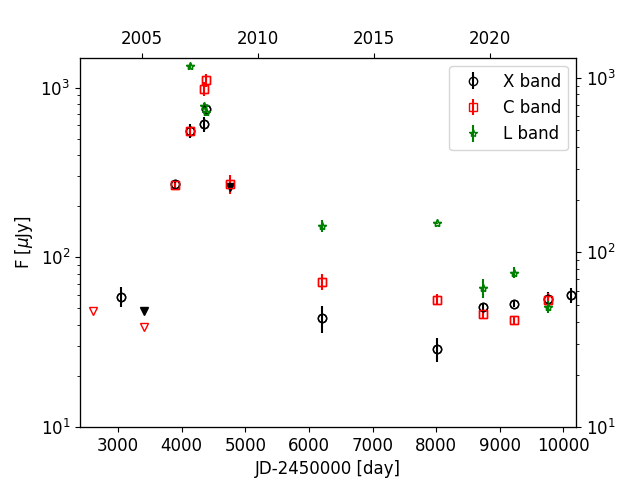}
\includegraphics[width=\columnwidth]{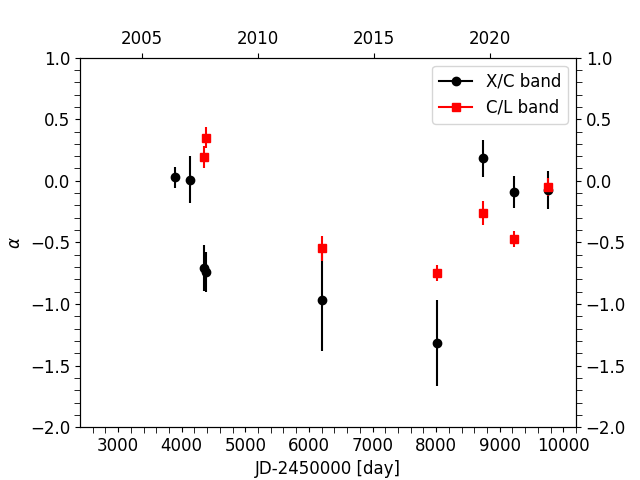}
\caption{ Top: evolution of radio continuum flux density of Sakurai's Object {in the X, C, and L bands}. Reverse triangles mark  three-sigma upper limits. Bottom: evolution of the and X/C band and C/L band spectral indices. The errorbars mark one-sigma error.}\label{Fig:fluxes}
\end{figure}

The compact radio source re-appeared in 2006 (Fig.\,\ref{Fig:fluxes}). The flux density in 2007 continued to increase and was more than tenfold higher than in 2004. The 2008 observation yielded four times lower flux density compared to 2007 at 4.8\,GHz. It indicates that the second maximum occurred in 2007--2008.

In 2012 flux density returned to the 2004 level. Then it faded even further in 2017 in the X and C bands. The L band 2017 flux density is affected by the old PN. The errorbars are larger due to smaller flux density. Interestingly, the X/C band spectral index $\alpha$ decreased from $0.03 \pm 0.09$ in 2006 to $-1.31 \pm 0.36$ in 2017 and the C/L band index decreased from $0.19 \pm 0.09$ in 2007 to $-0.75 \pm 0.07$ in 2017 (Fig.\,\ref{Fig:fluxes}). 

The flux density intensified again in 2019. The X/C band spectral index suddenly increased from $-1.31$ to $0.18 \pm 0.15$ at this time, which suggests that freshly ionized electrons appeared. The flux density and spectral index decreased in 2021 and remained at the same level in 2022--2023. This indicates that we missed the third maximum, which must have occurred between 2017 and 2019. 

%The 2022 spectrum was flat between 1.4 and 10 GHz.

%The maximum was relatively sharp and followed by an exponential decay.

%Flux density slightly faded in 2021. The spectral index decreased at similar rate to 2006--2007 observations.

%Finally, flux density slightly increased in 2022. The spectral index increased as well and remains almost flat in the whole observed $1.4-10\,\mathrm{GHz}$ range.

%\citet{2015ApJ...806..105S} observed change of the spectral index but reverse compared to our observations. The spectral index increased from $-0.54$ to $-0.28$ in IRAS\,15103–5754. In our case, spectral index decreased. 

%The radio flux of V605\,Aql increased between 1985 and 2005, followed by possible slight decrease in 2012. We scaled radio fluxes of V605\,Aql to the distance of Sakurai's Object and plotted them with fluxes of Sakurai's Object in Fig.\,2. Observations of V605\,Aql are shifted forward by 70 years, which corresponds to the difference of the dates of the outburtst. This can reflect the flux of Sakurai's Object in the future. A dashed vertical bar in Fig.\,2 corresponds to the date when the [WC]-type spectrum was observed by \cite{Seitter}.

%\begin{figure}
%\includegraphics[width=\columnwidth]{Images/saksi.png}
%\caption{Evolution of the radio flux density and 4.86/8.46 and %6/10\,GHz spectral index in Sakurai's Object.}\label{Fig:si}
%\end{figure}

\begin{figure}
\includegraphics[width=\columnwidth]{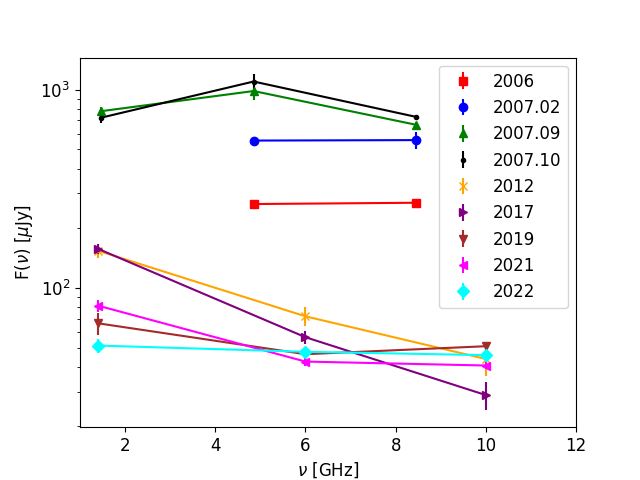}
\caption{Radio spectrum of Sakurai's Object between 2006 and 2023.}\label{Fig:sed}
\end{figure}

%In order to well constrain thermal emission, you need to measure both size and turnover.

\section{Discussion}

Non-thermal emission is observed in proto-planetary nebulae. It is characterized by a variability timescale of the order of months and negative spectral indices. \citet{2006MNRAS.369..189C} detected non-thermal emission with a spectral index of $\alpha = -0.8$ in OH\,354.88$-$0.54, due to shocked interaction between the fast stellar wind and nebular shell. IRAS\,15103$-$5754 intermittently showed non-thermal emission with varying flux density and spectral index \citep{2015ApJ...806..105S}. \citet{2009MNRAS.397.1386B} observed non-thermal emission in three proto-planetary nebulae. \citet{2017MNRAS.468.3450C} monitored non-thermal emission in five proto-planetary nebulae, explained by magnetic fields induced in shocks due to interacting winds, analogous to the interacting winds in binary stars with WR components. \citet{2024arXiv240603219O} discovered a non-thermal jet in the proto-PN IRAS\,18443-0231.

During the PN phase synchrotron emission is either not present or screened/suppressed by the thermal emission from the nebular shell \citep{2018MNRAS.479.5657H}. Spectral indices ranges from $-0.1$ for optically thin to $+2.0$ for optically thick frequency domain of the spectrum. Radio fluxes evolve on a timescale comparable to the evolutionary timescale of the central star or/and nebular expansion, which is typically too long to be observed. However, \citet{2008ApJ...681.1296Z} managed to observe slow evolution of radio flux in NGC\,7027 due to the expansion of the nebula and decrease of the number of ionizing photons from the central star. 

%\citet{1998ApJ...496..274M} observed chanes in radio images of IC\,4997 attributed to wind variability.

The emission observed in Sakurai's Object in 2004 was interpreted as thermal emission from freshly ionized ejecta by the central star \citep{2005Sci...308..231H}. In such case, optical and radio fluxes should increase with the increasing temperature and number of ionizing photons from the star, expected from evolutionary models. However, the observed decrease in the optical emission lines instead later indicated a shock origin \citep{2007A&A...471L...9V}. The decrease of the radio flux density after the 2007/2008 maximum ruled out photoionization by the central star.

The negative X/C band spectral index in 2006 and 2007 suggests synchrotron radiation from shocks (Figs\,\ref{Fig:fluxes},\,\ref{Fig:sed}). The spectral index and flux decreased between 2007 and 2017. It may suggest a decreasing contribution from free-free emission in the shock-ionized ejecta, which flattened the observed spectral index close to the maximum. Alternatively, it can suggest a loss of the energy of the electrons in the shock, causing a steepening of the radio spectrum. 

Radio emission from Sakurai's Object between 2004 and 2017 showed similar properties to proto-planetary nebulae. It has been non-thermal and highly variable. We attribute the observed maxima in the radio light curve to shocks induced by discrete mass ejection events. The photoionized shell, if present, had a much smaller contribution to the flux density than the non-thermal component. The timescale of the mass-loss variations is much shorter than the evolutionary timescale of the central star.

%The optical evolution of the newly ejected H-poor nebula is different from radio continuum emission. [N\,{\sc ii}] 6583\,\AA\ flux experienced a minimum in 2007 and started to gradually increase since then. After a long observing gap, the source shows a sudden increase of [O\,{\sc iii}] 5007\,\AA\ line by a factor $>20$ between 2021 and 2022 (PvH, private communication). Optical emission must originate from the outermost regions of the ejecta due to high extinction toward the central star. It could also be reflected from the far side of the dusty disk surrounding the central star. 

Sakurai's Object bears some resemblance to young stellar objects. Their radio spectrum consists of a few components. The stellar jet has a positive spectral index ($+0.6$), while distant lobes show a synchrotron component \citep[an average spectral index of $-0.55$ determined by][]{2016MNRAS.460.1039P}. Similar behavior to an outburst observed in 2006--2007 in Sakurai's Object (increasing flux and decreasing spectral index) was recorded for a protostar by \citet{2021MNRAS.501.5197O} and interpreted as an expanding bubble of gas. A similar effect is observed in supernovae outbursts \citep{2014ApJ...782...30Y}.

Interestingly, the maxima of the radio light curve occurred close to dust formation episodes observed in the infrared observations of Sakurai's Object. \citet{2004MNRAS.353L..41E} reported an increasing mass-loss rate between 2001 and 2003 marked by a dramatic increase in flux density at 450 and 850\,$\mathrm \mu m$, which may be related to the first maximum in radio between 2002 and 2004. A mass-loss event in 2008 was reported by \citet{2020MNRAS.493.1277E}. The 2007/2008 maximum in radio flux could correspond to this mass-loss episode. Another mass loss event was reported by \citet{2022MNRAS.511..713E} between 2013 and 2020. This could correspond to the increase of the flux density between 2017 and 2019.

Radio emission of Sakurai's Object between 2019 and 2023 became less variable and could be explained by thermal emission. In 2019 the spectral index reached a value of $+0.18 \pm 0.15$. In 2021--2022 the spectral index was close to zero. The source became resolved in 2021. This suggests low brightness temperature and thermal emission. The radio light curve does not show a decline after the 2017--2019 maximum, in contrast to the behavior observed after the 2007/2008 and the 2002--2004 maxima. It suggests that the ionized material is not recombining and that a stable source of ionizing photons may exist.

%or that an increasing contribution from the material photoionised by the central star is balanced by decreasing shock-ionised emission.

%Such discrete mass ejection events would produce patchy nebulosities.

%Indeed, patchy, hydrogen-deficient nebulae have been observed at other objects which experienced a previous VLTP event in the past (e.g. A\,58).
%The morphology of new ejecta stands in striking contrast to roundish, regular old PNe \citep{2008ASPC..391..177K}.

%\citet{1998ApJ...499L..83D} observed? non-thermal emission in A\,30.
%The value typical for thermal bi-conical jets is 0.6 \citep{1986ApJ...304..713R}. Thus, the jet, if existed about 2019, was (partially) optically thin. 

%Interaction of the ejecta caused shocks and observed radio emission. 

\begin{figure}
\includegraphics[width=\columnwidth]{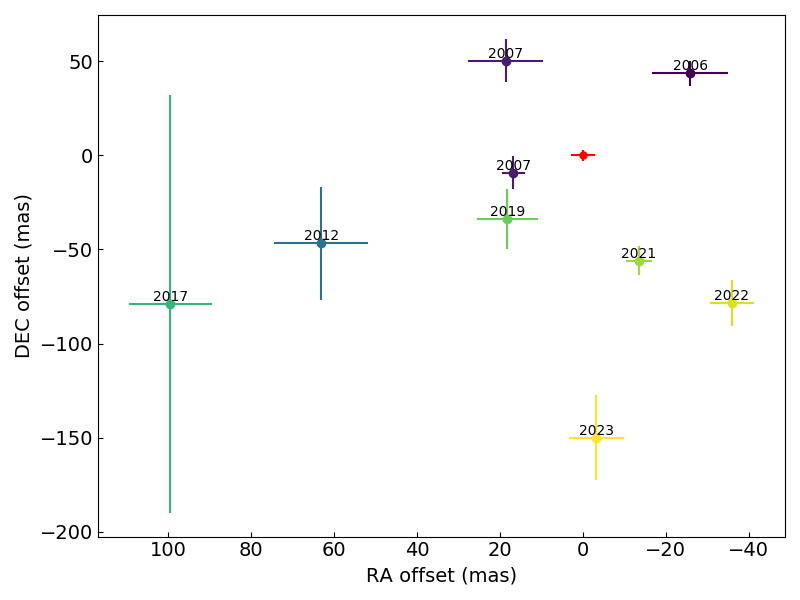}
\caption{Positions of ratio emission from the VLA X-band observations relative to the position of the 233\,GHz radio emission from \citet{2023A&A...677L...8T} marked with the red symbol. The 2004 year observation is ignored due to large errors.}\label{Fig:position}
\end{figure}

The position of the radio source was changing. Radio emission during the strongest maximum in 2007 was centered at the east to the presumed position of the central star, inside the disk seen at 233\,GHz by \citet{2023A&A...677L...8T} (Fig.\,\ref{Fig:position}). The projected distance between the position of the source in September and October 2007 would require a tangential velocity of the order of 10\,000\,km\,s$^{-1}$, which was not observed in Sakurai's Object. Thus, we conclude that different regions contributed to the radio emission at different times. 

In 2012 and 2017 the emission appeared to originate further away from the central star, where the density was lower and shock recombination took longer. 

In 2019--2023 the emission originated from a different region than in 2004--2017 and was centered south of the central star. The change of the position of the radio source between 2019 and 2023 corresponds to a tangential velocity of 440\,km/s at a distance of 3\,kpc. It may indicate an increasing contribution of emission from more distant regions or expansion.

The emission has been resolved since 2021. The 2023 image showed the largest extent (deconvolved full width at half maximum (FWHM)) of 0.40\,\arcsec\ in 2023 compared to 0.28\,\arcsec\ in 2021 and 0.20\,\arcsec\ in 2022. However, the uncertainties are too large to determine the expansion rate precisely.

%The true angular diameter of the observed source requires taking into account conversion factor, which depends on the beam size and intrinsic brightness distribution of a source \citep{2000MNRAS.314...99V}.

%We do not have unbiased information on the intrinsic size and brightness distribution of the source. This will become available as soon as the size of the emission becomes significantly larger than the synthesized beam. 

%(deconvolved FWHM of $0.40 \times 0.18$ and PA of $6$ deg). 

\citet{2023A&A...677L...8T} have observed a $\mathrm{H^{12}CN}$ bipolar emission extending more than 0.1\,\arcsec\ from the star and an expanding disk. The radio source appears to be aligned with the bipolar outflow and perpendicular to the dusty disk. The radial velocities of the $\mathrm{H^{12}CN}$ are between $-350$ and $300$ km/s. Radio continuum emission may be the counterpart of this structure. 

%Assuming an inclination angle of 18 deg, the kinematic age of this structure would be around 4 years.

The extent of the radio emission appears to be more compact than the separation of the expanding blobs (0.6\,\arcsec) previously reported by \citet{2020ApJ...904...34H}.
%separation of the expanding blobs observed by \citet{2020ApJ...904...34H} of 0.6\,arcsec in 2019. 
However, radio emission may originate from the inner part of these blobs facing the central star. The center of the radio continuum emission observed in 2019--2023 is moving to a position south of the central star (Fig.\,\ref{Fig:position}). This may indicate that southern blob is exposed to the radiation from the central star.

Radio images do not show the structure of the source in detail. It may contain an unresolved and resolved component. Thus, the vicinity of the central star may also contribute to radio emission. The inner part of the disk obscuring the central star may be photoionized by the central star.

It is not clear, whether the episodes of mass ejection observed in radio are related to the mass-loss from the star itself or to the accretion processes in
% the disk. The latter option would favour binary evolution.
a disk/torus structure around the central star. The latter option would favor a binary evolution scenario. 

%Evolutionary models of the VLTP do not predict distinct mass-loss episodes.

\section{Conclusions}

The VLA monitoring of Sakurai's Object allowed us to characterize an evolution of a radio emission of a VLTP object. Sakurai's Object showed non-thermal, highly variable emission in 2004-2017 and increasing contribution of thermal emission in 2019-2023. We attribute the observed evolution to intermittent shocks due to discrete mass loss events. The steepening spectral index between 2006 and 2017 indicates the diminishing contribution from the thermal component possibly due to recombination occurring in the shocks. Alternatively, it can indicate decreasing energy of the electrons in the shock. Further monitoring will unambiguously reveal thermal radio emission due to the photoionization of the ejecta by the reheating central star. Our observations can help us to identify VLTP events, expected to occur at the rate of one per decade in our Galaxy, as transients in high-sensitivity radio surveys.

\begin{acknowledgements}
The National Radio Astronomy Observatory is a facility of the National Science Foundation operated under cooperative agreement by Associated Universities, Inc. The Australia Telescope Compact Array is part of the Australia Telescope National Facility (https://ror.org/05qajvd42) which is funded by the Australian Government for operation as a National Facility managed by CSIRO. We acknowledge the Gomeroi people as the Traditional Owners of the Observatory site. Computations were carried out using the computers of Centre of Informatics Tricity Academic Supercomputer \& Network. D.B. acknowledges support by Fondecyt grant 11230261. 
\end{acknowledgements}


\bibliographystyle{aa}
\bibliography{sakurai}

\onecolumn

\begin{appendix}

%\longtab{
\begin{landscape}

\section{Details of the radio observations of Sakurai's Object.}

\begin{longtable}{ccccccccccc}
\caption{\label{tab:observations} Radio observations of Sakurai's Object. The columns indicate date of observation, central frequency, beam size and position angle, deconvolved FWHMs and position angle, peak and integrated flux density and source coordinates. U in deconvolved size column stands for unresolved source.} \\
\hline\hline
Date & Freq. [GHz] & Beam size & PA [deg] & Size [mas] & PA [deg] & & Fpeak [${\mu}\mathrm{Jy}$] & Fint [${\mu}\mathrm{Jy}$]& RA & DEC \\
\hline
\endfirsthead
\caption{continued.}\\
\hline\hline
Date & Freq. [GHz] & Beam size & PA [deg] & Size [mas] & PA [deg] & & Fpeak [${\mu}\mathrm{Jy}$] & Fint [${\mu}\mathrm{Jy}$]& RA & DEC \\
\hline
\endhead
\hline
\endfoot
\\[-1.5ex]
%2002-11-29 & 4.86 & $6.43 \times 3.93$ & $1.48$ &   & & & $<48$ &  &  &  \\
2004-02-05 & 8.46 & $3.71 \times 2.13$ & $4.56$ & U & & & $58.7 \pm 8.0$ & $60 \pm 15$ &  17:52:32.6860(76) & $-17$:41:07.1055(0.2692) \\
2005-02-05 & 8.46 & $1.37 \times 0.97$ & $67.55$ &  & & & $<48$ &  &  &  \\
2005-02-04 & 4.86 & $3.71 \times 2.13$ & $4.56$ &  & & & $<39$ &  &  &  \\
2006-06-11 & 8.46 & $0.74 \times 0.63$ & $56.38$ & U & & & $269.0 \pm 7.3$ & $258 \pm 12$ &  17:52:32.69718(65) & $-17$:41:07.87142(667) \\
2006-06-12 & 4.86 & $1.41 \times 1.05$ & $-4.93$ & U & & & $265 \pm 10$ & $304 \pm 20$ &  17:52:32.6989(0.0011) & $-17$:41:07.9316(0.0310) \\
2007-02-02 & 8.46 & $7.04 \times 4.27$ & $86.38$ & U & & & $557 \pm 53$ & $402 \pm 24$ &  17:52:32.687(15) & $-17$:41:08.354(126) \\
2007-02-02 & 4.86 & $12.85 \times 7.54$ & $81.17$ & U & & & $554 \pm 26$ & $840 \pm 61$ &  17:52:32.625(25) & $-17$:41:08.584(171) \\
2007-02-02 & 1.46 & $21.2 \times 6.0$ & $-78.70$ & U & & & $1335 \pm 33$ & $1509 \pm 65$ &  17:52:32.281(48) & $-17$:41:06.433(253) \\
2007-09-04/05 & 8.46 & $0.39 \times 0.21$ & $-15.84$ & U & & & $665 \pm 26$ & $609 \pm 49$ &  17:52:32.70018(19) & $-17$:41:07.92426(875) \\
2007-09-04/05 & 4.86 & $0.63 \times 0.32$ & $172.4$ & U  & & & $985 \pm 95$ & $863 \pm 48$ &  17:52:32.69839(48) & $-17$:41:07.91452(1992) \\
2007-09-04/05 & 1.46 & $2.57 \times 1.61$ & $-26.37$ & U  & & & $779 \pm 35$& $750 \pm 63$ &  17:52:32.8383(58) & $-17$:41:06.0416(1586) \\
2007-10-09 & 8.46 & $1.06 \times 0.63$ & $30.69$ & U & & & $729 \pm 17$ & $745 \pm 31$ &  17:52:32.70031(64) & $-17$:41:07.86463(1162) \\
2007-10-09 & 4.86 & $1.65 \times 1.08$ & $24.44$ & U & & & $1100 \pm 95$ & $858 \pm 145$ &  17:52:32.7097(33) & $-17$:41:07.9216(547) \\
2007-10-09 & 1.46 & $4.55 \times 2.28$ & $62.93$ & U & & & $722 \pm 40$ & $693 \pm 73$ &  17:52:32.6802(96) & $-17$:41:07.7624(0.0657) \\
%2007-10-09 &  &  &  &  & & &  &  &  17:52:32.7097(33) & $-17$:41:07.9216(547) \\
A2008-10-27 & 8.46 & $3.47 \times 0.95$ & 2.16 &  & & & $<260$ &  &   &  \\
A2008-10-27 & 4.8 & $6.36 \times 1.70$ & 1.61 & U & & & $273 \pm 34$ & $232 \pm 65$ & 17:52:32.6840(42) & $-17$:41:07.4502(4730) \\
2012-10-08 & 9 & $0.46 \times 0.23$ & 16.85 & U & & & $43.8 \pm 7.8$ & $43.3 \pm 39$ & 17:52:32.70344(80) & $-17$:41:07.96184(2997) \\
2012-10-11 & 5.5 & $0.58 \times 0.31$ & 22.72 & U & & & $71.9 \pm 8.0$ & $80 \pm 16$ & 17:52:32.7002(11) & $-17$:41:07.7929(378) \\
2012-10-11 & 1.5 & $2.6 \times 1.31$ & 38.89 & U & & & $154 \pm 13$ & $133 \pm 27$ & 17:52:32.6827(104) & $-17$:41:08.3512(1845) \\
%2013-01-12 & 10 & & & & & & & &  \\
2017-09-12    & 10 & $1.54 \times 0.71$ & $-31.12$ & U & & & $28.8 \pm 4.7$ & $22.1 \pm 7.9$ &17:52:32.7060(70) & $-17$:41:07.994(111) \\
2017-09-12    & 6 & $2.33 \times 1.17$ & $-25.29$& U  & & & $56.4 \pm 4.2$ & $86.0 \pm 10.0$ &17:52:32.7055(50) & $-17$:41:08.021(91) \\
2017-09-12    & 1.52 & $7.75 \times 4.14$ & $-23.11$&   U & & & $158.8 \pm 0.7$ & $324.7 \pm 2$ & 17:52:32.6665(102) & $-17$:41:09.1266(3602)   \\
2019-09-04    & 10 & $0.39 \times 0.21$ & $20.32$ & U & & & $50.8 \pm 2.8$ & $64.7 \pm 6.2$ &17:52:32.70028(51) & $-17$:41:07.949(16) \\
2019-09-04    & 6 & $0.73 \times 0.36$ & $31.12$& U & & & $46.3 \pm 2.4$ & $47.9 \pm 4.4$ &17:52:32.70028(74) & $-17$:41:07.96750(1563) \\
2019-09-04    & 1.52 & $3.92 \times 1.36$ & 41.26 & U &  & & $66.3 \pm 8.5$ & $86 \pm 19$ &17:52:32.701(17) & $-17$:41:07.929(253) \\
2021-01-09    & 10 & $0.40 \times 0.22$ & $19.24$ & $0.28 \times 0.09$ & $16.4 \pm 5.8$& & $40.6 \pm 1.5$ & $53.4 \pm 2.5$ & 17:52:32.69805(22) & $-17$:41:07.97108(766) \\
2021-01-09    & 6 & $0.77 \times 0.38$ & $31.76$ & U & & & $42.5 \pm 2.3$ &  $41.9 \pm 4.3$ &17:52:32.69921(85) & $-17$:41:07.91208(2484) \\
2021-01-09    & 1.52 & $4.15 \times 1.42$ & $40.44$ & U & & & $81.4 \pm 5.8$ &  $153 \pm 16$ &17:52:32.6880(74) & $-17$:41:08.1036(1219) \\
2022-06-25    & 10 & $0.32 \times 0.18$ & $18.46$& $0.20 \times 0.06$&  $18 \pm 18$& &$45.8 \pm 2.7$ & $56.8 \pm 5.8$ &17:52:32.69647(37) & $-17$:41:07.99331(1231) \\
2022-06-25    & 6 & $0.57 \times 0.28$ & $27.11$ & $0.192 \times 0.08$& $18 \pm 12$ & & $47.6 \pm 2.6$ & $56.3 \pm 5.4$ &17:52:32.69623(46) & $-17$:41:08.005(15) \\
2022-06-25    & 1.52 & $2.63 \times 1.08$ & $35.18$& U& & & $51.2 \pm 4.2$ & $84 \pm 11$  &17:52:32.7722(78) & $-17$:41:08.0758(0635) \\
2023-06-23    & 10 & $0.43 \times 0.20$ & $-11.49$ & $0.40 \times 0.18$& $6 \pm 16$ & & $32.1 \pm 2.3$ & $60.0 \pm 6.4$ &17:52:32.69877(46) & $-17$:41:08.06499(2278)
\end{longtable}
\end{landscape}
\end{appendix}

\end{document}